\documentclass[useAMS,usenatbib]{mn2e}
\usepackage{graphicx}

\def\ergs{${\rm erg\,cm^{-2}\,s^{-1}}$ }
\def\ergse{${\rm erg\,cm^{-2}\,s^{-1}}$}

\def\ucre{${\rm cts\,s^{-1}}$}
\def\ulum{${\rm erg\,s^{-1}}$ }
\def\ulume{${\rm erg\,s^{-1}}$}
\def\nh{$N_{\rm H}$ }
\def\nhe{$N_{\rm H}$}

\def\unh{${\rm cm^{-2}}$ }

\def\gnh{$N_{\rm H}^{\rm Gal}$ }
\def\gnhe{$N_{\rm H}^{\rm Gal}$}
\def\ngc{NGC\,7589 }
\def\ngce{NGC\,7589}

\def\eddrat{$L_{\rm bol}/L_{\rm Edd}$ }
\def\eddrate{$L_{\rm bol}/L_{\rm Edd}$}
\def\msun{M$_{\sun}$ } 
\def\msune{M$_{\sun}$} 

\title[NGC\,7589]
{Discovery of high-amplitude X-ray variability in the Seyfert--LINER
    transition galaxy NGC\,7589}
\author[W. Yuan, et al.]{W. Yuan$^{1}$\thanks{E-mail:
wmy@ast.cam.ac.uk (wy)},
St. Komossa$^{2}$,
D. Xu$^{3}$, 
J.P. Osborne$^{4}$, 
M.G. Watson$^{4}$, 
R.G. McMahon$^{1}$\\
$^{1}$ University of Cambridge, Institute of Astronomy, 
Madingley Road, Cambridge, CB3 0HA\\
$^{2}$ Max-Planck Institut f\"ur Extraterrestriche Physik,
        Postfach 1312, 85741 Garching, Germany\\
$^{3}$ National Astronomical Observatories, Chinese Academy of
Sciences, Beijing 100012, China\\
$^{4}$ Department of Physics and Astronomy, University of Leicester, LE1 7RH
}

\begin{document}

\date{Accepted 2004 July 29, Received 2004 June 13}

\pagerange{\pageref{firstpage}--\pageref{lastpage}} \pubyear{2003}

\maketitle

\label{firstpage}

\begin{abstract}
We present the first result of a programme to search for large
flux variations in the X-ray sources of the XMM Serendipitous Survey
compared to previous ROSAT observations.
An increase in X-ray flux  by a factor $>$10
was discovered from the nucleus of the galaxy NGC\,7589
on a timescale of less than 5\,years.
The 0.4--10\,keV XMM spectrum can be approximated by a power-law with
photon index of 1.7--1.8, though
it seems to flatten above 5\,keV, 
suggesting a possible complex model, 
such as partial covering or disc reflection.
A classification based on an analysis of its optical spectrum
places \ngc in the Seyfert region, but
close to the Seyfert--LINER border-line on the AGN diagnostic diagrams.
We classify \ngc as either Seyfert\,1.9 or LINER\,I,
in the light of the detection of a broad H$\alpha$ line,
which makes \ngc an AGN in the low-luminosity regime.
We interpret the observed variability in terms of 
either changes in covering factor of absorbing gas in the AGN,
or variability in the intrinsic X-ray luminosity.
Should the latter be the case, 
the inferred Eddington accretion rate increased from 
the  radiatively inefficient accretion dominated regime to 
a value close to the putative critical value, 
at which a transition of the accretion mode is supposed to take place.
This possibility presents a new prospect of studying 
accretion physics in the central black holes of external galaxies 
by direct observing
changes of `spectral state', as is common
in stellar black hole binary systems.
\end{abstract}

\begin{keywords}
galaxies: active galaxies: individual: NGC\,7589 - X-ray: galaxies
\end{keywords}

\section{Introduction}
In contrast to typical active galactic nuclei (AGN),
low luminosity AGNs (LLAGNs) as a class
radiate at a power much lower than the Eddington luminosity\footnote
{$L_{\rm Edd}$=1.26$M_{\rm bh}10^{38}$\,\ulume, $M_{\rm bh}$ 
is black hole mass in \msune.},
i.e.\ in terms of the bolometric/Eddington luminosity ratio
(Eddington ratio), \eddrate$<$\,0.01 (Ho 2004).
As such, black holes in LLAGNs  are hypothesised to accrete via 
radiatively inefficient accretion flows 
(RIAF, see Quataert 2001 for a review), 
such as an advection-dominated accretion flow
(ADAF, see Narayan et al.\ 1998 for a review),
rather than an optically thick, geometrically thin
standard accretion disc (thin disc hereafter, Shakura \& Sunyaev 1973).
In fact, the high-end of their \eddrat distribution encompasses the
critical value\footnote
{Hypothesised as 0.01--0.1 Eddington ratio for ADAF (Narayan et al.\ 1998); 
however, observations show that the transition always occurs at a
luminosity around  $10^{37}$\,\ulum (e.g.\ Tanaka 1999), 
which corresponds to a few per cent Eddington ratio.}
above which a RIAF is to be replaced with a standard thin disc.
As a consequence, in LLAGNs exhibiting violent variability of luminosity,
a transition of the accretion mode is expected to take place
once \eddrate crosses the critical value.
This is where observations can be used to test black hole
accretion theories.
While such a transition of the accretion mode
can, indeed, explain the observed changes of the `spectral
states' in X-ray binaries 
(e.g.\ Esin et al.\ 1997, Meyer et al.\ 2000),
the situation is not clear in massive extragalactic black hole systems.
On galactic scales the major difficulty in observations 
is the long timescales of the X-ray variability, 
which is roughly scaled with the black hole mass.
Timely detection of violent variability in AGN,
especially in LLAGN, is of particular interest with this regard.

About 40 per cent of nearby galaxies exhibit LLAGN activity
(Ho et al.\ 1997), the majority among of 
belong to a class known as
low-ionisation nuclear emission-line regions (LINER, Heckman 1980).
Large amplitude X-ray variability is
sometimes seen in low luminosity  Seyferts (e.g.\ Nandra et al.\ 1997),
whereas no similar behaviour is found in LINERs 
(e.g.\ Ptak et al.\ 1998, Komossa et al.\ 1999, Terashima et al.\ 2002).

In this letter we report the discovery of large amplitude 
X-ray variability in the galaxy \ngc
at a redshift of 0.0298.
This is one of the first results
of a  search of long-term highly variable X-ray  sources
(in preparation; see also Yuan et al.\ 2002)
of the XMM Serendipitous Survey (Watson et al.\ 2002).
We selected candidates which varied by a factor of at least 10 for
further follow-up studies in the X-ray and other wavebands.
We used a luminosity distance of 123.3\,Mpc calculated from
the radial velocity of 8562\,km\,s$^{-1}$ (LEDA)
relative to the cosmic microwave background 
by assuming  $H_0$=71\,km\,s$^{-1}$\,Mpc$^{-1}$,
$\Omega_{\Lambda}$=0.73, and $\Omega_{\rm m}$=0.27.

\section[]{The XMM X-ray data of \ngc}
\label{sect:xmm}
\subsection{XMM observations}
\begin{figure}
\includegraphics[angle=0,width=\hsize]{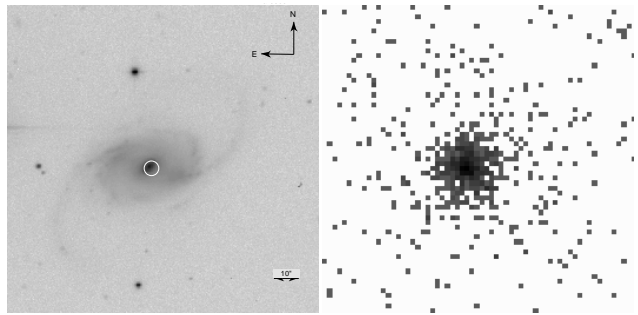}
 \caption{\label{fig:xmm}
     Optical r'-band (left) and X-ray (right, XMM MOS-1, orbit 272) 
    images of \ngc on the same scales.
	A circle on the optical image represents the position and 
        its 3$\sigma$ error of the X-ray source.
	A 10\arcsec angular distance corresponds to 5.65\,kpc.
      }
\end{figure}

The XMM observations, data screening, and source extraction 
are summarised in Table\,1.
It should be noted that the PN camera was not active during 
the first observation in the XMM orbit 272.
The XMM Science Analysis System (SAS, v.5.4) 
was used for data reduction.
Source X-ray events
were extracted from a circle of 32\,\arcsec radius.
Background events were extracted from
source-free regions using a concentric annulus of 40/120\,\arcsec
radii for the MOS detectors,
and circles of  32\,\arcsec radius at 
the same CCD read-out column as the source position for the PN detector.
   \begin{table}
      \caption[]{Summary of the XMM observations and data reduction}
         \label{tab:xmmobs}
         \begin{tabular}{lccc}
            \hline \hline
            \noalign{\smallskip}
    orbit number    & \multicolumn{1}{c}{orbit 272}  &
      \multicolumn{2}{c}{orbit 361}  \\	 
               \cline{3-4} \noalign{\smallskip}
     detectors       &  MOS1/2     &   MOS1/2   & PN \\
            \noalign{\smallskip}
            \hline
            \noalign{\smallskip}
date   & Jun. 03 2001& \multicolumn{2}{c}{Nov. 11 2001}\\
observation ID & 0066950301 & \multicolumn{2}{c}{0066950401}\\
duration (ks) & 12.2 & 12.6 & 13.2\\
GTI  CR$^a$& 0.35   & 0.35 & 1.0 \\
events pattern     & 0--12  & 0--12 & 0--4 \\
good exposure (s)    & 7366/7151 & 9175/9826 & 8133\\
net source CTS  & 555/503  & 243/276 & 843 \\
src CR ($10^{-2}$\ucre) &7.5/7.0$\pm0.4$  & 2.6/2.8$\pm0.3$ & 10.4 $\pm0.7$\\
            \hline \hline
         \end{tabular}
\begin{list}{}{}
\item[$^{\mathrm{a}}$] Count rate criteria for rejecting high-background 
periods identified by single events (PI$>10,000$, Pattern=0) 
in units of \ucre.
\end{list}
   \end{table}
In both observations an X-ray source was detected at the position
RA=23h 18m 15.6s, Dec=0$^o$ 15\arcmin 38.9\arcsec (J2000), 
coincident with the nucleus of \ngc
(Fig.\,\ref{fig:xmm}).
The source spatial profile in the 0.3--2keV band 
was found to be 
point-like.
No time variability was found on timescales shorter than 
the XMM observational intervals of about 12\,ksec.
The galaxy was also observed with the XMM Optical Monitor (OM)
with the UVW1 filter (2000--4000\AA)
and was detected as 
an extended source. 

\subsection[]{The XMM X-ray spectra and fluxes}
\label{sect:xspec}
The MOS1 and MOS2 spectra were co-added to 
produce a single, combined spectrum.
We used XSPEC (v.11.3) for spectral modelling.
Spectral bins below 0.4\,keV were excluded because of uncertain 
calibration below 0.35\,keV (XMM-SOC-CAL-TN-0018).
From orbit 361 we have both PN and MOS spectra; 
they were jointly fitted with independent normalisations.
In comparisons of the two observations,
the fitted spectral parameters agree with each other
for most or all the parameters except 
the normalisations, which differ by a factor of $\simeq2.5$.
To improve photon statistics,
we also performed 
fits to the two spectra 
from the different epochs
with most of the parameter values tied together.
The results are summarised in Table\,\ref{tab:spec_fit} and
explained in below.

\begin{figure}
   \includegraphics[angle=270,width=\hsize]{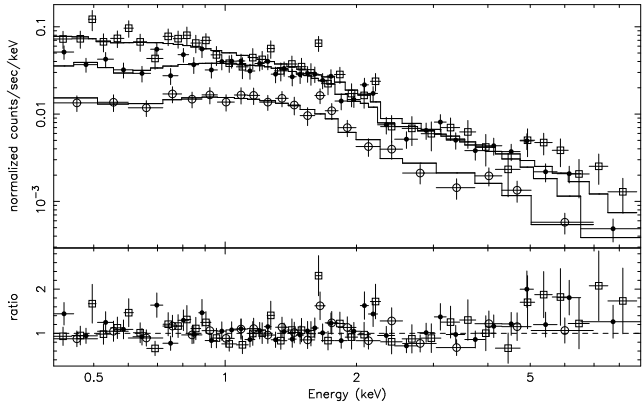}
      \caption{\label{fig:pl_extrap}
     XMM spectral data, model fit and the residual as data-to-model ratio.
   The model is for  a joint power-law fit to the
	spectra of the combined MOS in orbit 272 (dots), 
     and the MOS (circles) and the PN (squares) in orbit 361,
       in the restricted 0.4--2\,keV band 
         and extrapolated to higher energies.
	A spectral flattening can be seen for both orbits.
      }
   \end{figure}

{\em Power-law models}:
A simple absorbed power law gives statistically acceptable fits to 
all the spectra; however, the fitted \nh is lower than 
the Galactic HI column density along the
line-of-sight \gnhe=4.06\,$10^{20}$\,\unh 
(at the 92 per cent confidence level).
The fits over the restricted 0.4--2\,keV and 2-10\,keV ranges
yielded $\Gamma_{\rm low-E}$=1.72$\pm$0.12 and 
$\Gamma_{\rm high-E}$=1.42$^{+0.20}_{-0.26}$, respectively,
indicating a spectral flattening towards high energies. 
This can be seen from the residuals in Fig.\,\ref{fig:pl_extrap}, 
which compares the data to a fitted model using  
only the 0.4--2\,keV spectra.
In fact, a broken power-law model of
$\Gamma_{\rm low-E}$=1.75$^{+0.10}_{-0.16}$ and
$\Gamma_{\rm high-E}$=1.03$^{+0.40}_{-0.47}$
improved the fit significantly,
with $\Delta \chi^2=-10$ for 2 additional free parameters;
the  $F$-test gave a probability level for no improvement over power-law
as being $<$1 per cent.
The fitted \nh was also recovered to value consistent  with \gnhe.

{\em Complex spectral models}:
Spectral flattening is often seen in AGNs as  a result of
scattered X-rays, likely by an accretion disc, 
or X-rays transmitted 
through optically thick gas.
Accordingly, we fit disc reflection  
and partial covering models to the spectra.
Both types of models, either cold or ionised, 
were found to provide excellent and statistically
indistinguishable fits to all the data.
The overall absorption \nh values were in general consistent with \gnh.
The disc inclination was not constrained and 
so was fixed at 30\,degrees.
The disc reflection models require a larger reflection/direct ratio
(i.e.\ $>2$) than that from a medium subtending a $4\pi$ solid angle, 
implying that the primary continuum is partially obscured in this case.
No significant  iron K$\alpha$ line  was seen at around 6.4\,keV;
90 per cent upper limits for the line equivalent width were found to be
550\,eV and 110\,eV for the 1st and 2nd observations, respectively.
Modelling the high-energy excess simply by an
extremely broadened iron line (the {\it diskline} model) 
gave no satisfactory fit.

\begin{table}
\caption{Spectral fits to XMM data for \ngc}
\label{tab:spec_fit}
\begin{tabular}{lllc} \hline\hline
data set     & orbit 272 &  orbit 361 & \multicolumn{1}{c}{both orbits} \\
detectors    & MOS       & MOS+PN &  orb272/orb361\\ \hline
\multicolumn{4}{l}{power law, fixed \nh = \gnh =4.06\,$10^{20}$\,\unh}\\
$\Gamma$    & 1.69$\pm$0.09 & 1.68$\pm$0.08 & \multicolumn{1}{c}{1.68$\pm$0.06}\\
$N_{\rm 1keV}$\,/$10^{-5}$   & 14.2$\pm$0.9 & 5.5$\pm$0.5& 14.1$\pm$1.0 /5.4$\pm$0.5 \\
$\chi^2$/dof& 44/39      &  58/62     & \multicolumn{1}{c}{103/102} \\ \hline
\multicolumn{4}{l}{partial covering by cold matter, \nh = \gnh} \\
$\Gamma$     & 1.80$^{+0.07}_{-0.11}$& 1.74$\pm0.10$         & 1.77$^{+0.04}_{-0.08}$ \\
\nh /$10^{22}$& 30$^{+85}_{-20}$      & 48$^{+68}_{-37}$      & 40$^{+62}_{-23}$ \\
covering factor & 0.54$^{+0.35}_{-0.27}$& 0.54$^{+0.46}_{-0.24}$& 0.55$^{+0.25}_{-0.25}$ \\
$\chi^2$/dof & 38/37                 & 55/60                 & 93/100 \\ \hline
\multicolumn{4}{l}{partial covering by ionised absorber (absori), \nh=\gnh} \\
$\Gamma$      & 1.77$\pm0.14$ &  1.73$\pm0.10$ & 1.75$\pm0.08$ \\
\nh /$10^{22}$ & 70$_{-56}$ & 74$_{-61}$ & 76$_{-30}$  \\
$\xi$         & 123$^{+2755}_{-123}$ & 31 & 137$^{+468}_{-137}$/60$^{+166}_{-60}$ \\
covering factor  & 0.54$^{+0.26}_{-0.39}$ & 0.55$^{+0.28}_{-0.50}$ &0.53$^{+0.17}_{-0.25}$ \\
$\chi^2$/dof  & 37/35  &  55/58      & 92/96 \\ \hline%
\multicolumn{4}{l}{reflection from cold disc (pexrav), fixed \nh=4.06\,$10^{20}$\,\unh} \\
$\Gamma$ & 1.85$\pm0.14$      & 1.74$^{+0.15}_{-0.06}$ & 1.79$\pm0.08$ \\
rel-refl & 8.0$^{+4.6}_{-5.4}$ & 3.5$^{+5.0}_{-3.5}$ &5.9$^{+4.8}_{-3.9}$/5.0$^{+4.9}_{-4.0}$\\
$\chi^2$/dof  & 38/38          &  56/61                 & 95/100 \\ \hline
\multicolumn{4}{l}{reflection from ionised disc (pexriv), fixed \nh=\gnh} \\
$\Gamma$      & 1.77$\pm0.20$ &  1.73$\pm0.13$ & 1.75$^{+0.10}_{-0.08}$ \\
$\xi$         & 157 &  3  & 186$^{+224}_{-186}$/4$^{+217}_{-4}$ \\
rel-refl      & 3.5$^{+11.3}_{-2.9}$ & 3.3$^{+5.1}_{-3.3}$ & 3.2$^{+5.0}_{-2.4}$ \\ 
$\chi^2$/dof  & 37/37                &  56/60              & 94/99 \\ \hline\hline
\end{tabular}
Errors are at the 90 per cent level when quoted; 
if not given they are not
constrained within the physically meaningful ranges. \\
\nhe: column density in units of cm$^{-2}$ \\
$N_{1keV}$: normalization at 1\,keV in units of
phot\,cm$^{-2}$\,s$^{-1}$\,keV$^{-1}$; 
values quoted are for the combined MOS detectors.\\
rel-refl: factor of reflection  relative  to the primary continuum.\\
$\xi$: ionisation parameter defined as $L/nR^2$ 
for absorber with density $n$ at a distance $R$ to a source
with an ionising luminosity $L$
(Done et al.\ 1992).
\end{table}

{\em X-ray fluxes and luminosities}:
The X-ray fluxes  were calculated using the best-fit power-law
spectral model with \nhe=\gnhe, 
i.e.\ $\Gamma$=1.68\,(1.73) for 0.5--10\,(0.5-3)\,keV.
In the 0.5--10\,keV band, the Galactic absorption corrected 
fluxes are 9.1\,$10^{-13}$ and 3.7\,$10^{-13}$\,\ergse, 
for orbits 277 and 361 respectively, 
which correspond to  luminosities of 
1.7\,$10^{42}$\,\ulume and 0.7\,$10^{42}$\,\ulume, respectively.
In consideration of the likely presence of 
the partial covering or disc reflection 
(assuming a flat disc subtending a 2$\pi$ solid angle)
component, the intrinsic X-ray luminosities are likely
a factor of $\sim$2 and 3--5  higher, respectively.

\section{The long-term X-ray variability}
\label{sect:xvary}
\subsection{Fluxes from the ROSAT and Einstein data}
\ngc was within the field of view (10\,arcmin off-axis) 
    of a 5.4\,ksec pointing observation 
    made by ROSAT  with its PSPC detector (0.1--2.4 keV) on 16--17 June, 1995.
    No source was detected at the position of \ngce. 
    We derived a limit on the source flux in two ways. 
    The first was a simple estimate  using 
    the countrate of the weakest detected 
    source neighbouring \ngce.
This led to a conservative     upper limit on the source count rate 
of 2.7\,$\times 10^{-3}$\,\ucre.
    A more rigorous estimation was made using the PSPC X-ray  image.
    Photon counts from the background and a possible source
    were extracted from a circle of 45\,\arcsec radius 
    (enclosing 95 per cent power of the PSF) at the position of \ngce,
    yielding a measured C$_{\rm s+b}$=17\,cts.	
    Following Poisson statistics, the expected counts 
    	\={C}$_{\rm s+b}$ is distributed as \={C}$_{\rm s+b}\leq$24\,(27)
	at the (one-tail) probability of 95\,(99) per cent.
    The contribution of the background 
    was estimated as C$_{\rm b}$=9.5\,cts, 
    using the averaged local background estimated
    from an annulus of of 90\arcsec/225\arcsec radii.	
	The expected net source counts was estimated to be
	\={C}$_{\rm s}\le$ \={C}$_{\rm s+b}$-C$_{\rm b}$=14.5\,(17.5)\,cts.
    After correction for the vignetting and 
    extraction aperture, the source count rate 
	was estimated to be $\le$2.9\,(3.4)\,$10^{-3}$\,\ucre.
    Assuming the spectrum to be the same as that measured with XMM,     
    we derived the 0.1--2.4\,keV flux limit to be
    5.9\,(7.0)\,\,$10^{-14}$\,\ergs at the 
    95 (99) per cent level, after correction for the Galactic absorption.

    A less stringent flux limit of 5.6\,$10^{-13}$\,\ergs
     was set by the data extracted from the 
     ROSAT All-sky Survey performed in 1992.
By searching the data archives of various X-ray satellites,
we found that, in addition to ROSAT,
\ngc was also serendipitously observed by the {\it Einstein}
Observatory with its IPC detector 
at an off-axis angle of 10.8\,arcmin for 2540\,sec in May 1980.
Similar estimation using a source extraction radius of 3\arcmin,
yielded a limit on the
unabsorbed flux of
3.5\,$10^{-13}$\,\ergse.
Both limits are quoted at the 95 per cent level.

\subsection[]{Variability in the X-rays and UV}
\begin{figure}
\includegraphics[angle=0,width=\hsize, height=0.65\hsize]{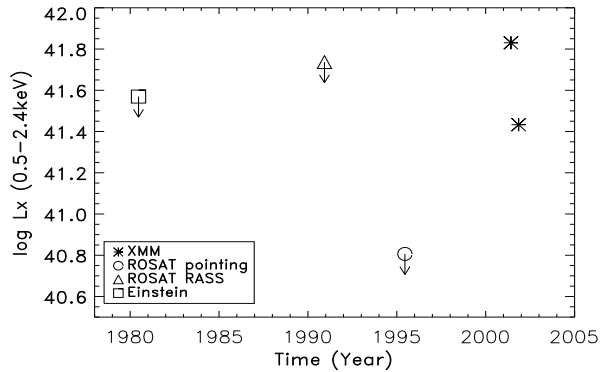}
 \caption{\label{fig:xlc}
     Long-term light curve of  0.5--2.4\,keV  X-ray luminosity of
 \ngce.  Upper limits  are at the 95 per cent confidence level.}
\end{figure}

For comparison of the fluxes among the different missions
we used the overlapping 0.5--2.4\,keV energy band.
The light curve in Fig.\,\ref{fig:xlc} reveals 
a dramatic increase in luminosity, by a factor of $>$\,10,
occurring sometime between the ROSAT pointing and the XMM
observation in orbit 272, on a timescale $\la$5\,years.
Given the poor sampling, the flux seen by XMM may not represent 
the flux at the `highest state', which means that the
peak luminosity could be even higher.
Therefore the actual variability amplitude must be $>$10.
The flux decreased by a factor of 2.5 
between
the two XMM observations
over a period of 5\,months,
whereas the overall 0.4--10\,keV spectral shape 
appeared to remain unchanged.

The OM UV magnitude within a central 6\arcsec-radius aperture
was measured to be 16.99$\pm$0.06 
and 17.03$\pm$0.05 for the orbit 272 and 361, respectively.
However, the emission is dominated by the extended bulge and
no significant point-like source is present;
thus no meaningful constraint on  variability  could be obtained.

\section[]{Optical spectroscopic classification}
\label{sect:class}
\ngc was observed spectroscopically in the Sloan Digital Sky Survey
(SDSS; York et al.\ 2000) on Sep.\ 29th, 2000,
about 8 months prior to the high state caught by XMM.
The spectrum was taken from the SDSS Data Release 2.
The contribution of stellar light from the host galaxy was
subtracted from the observed spectrum
by modelling stellar absorption lines
employing a range of galaxy spectral templates,
using the method developed by Li et al.\ (2004)
(A full account of the spectral analysis and estimation of the
black hole mass for \ngc is to be presented elsewhere.)
Single-component Gaussians were fit to the emission lines.
This provided satisfactory fits, except for
H$\alpha$ which required the presence of a second, broad component.
The broad H$\alpha$ line, which was blueshifted, 
had a width of FWHM
$\approx$3428\,km\,s$^{-1}$,
compared to the narrow lines
of FWHM $\approx$228\,km\,s$^{-1}$.

\begin{figure}
 \includegraphics[angle=0,width=\hsize]{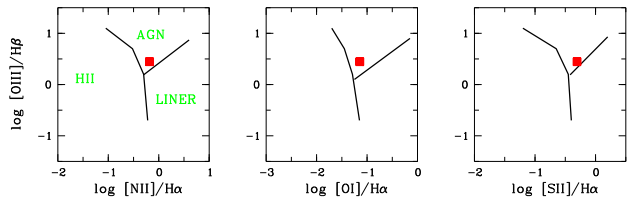}
 \caption{\label{fig:optdiagn}
   Position of \ngc in the
   diagnostics diagrams based on its optical emission line ratios. 
      }
\end{figure}

The emission-line ratios
place NGC\,7589 in the Seyfert region of the diagnostic diagrams
of Veilleux \& Osterbrock (1987), close to the border-line 
with the LINERs (see Fig.\,\ref{fig:optdiagn}).
The same conclusion also holds
if we use the classical
definition of LINERs (Heckman et al. 1980) which only
makes use of Oxygen emission-line ratios.
For LINERs,
[OII]$\lambda$3727/[OIII]$\lambda$5007$>$1 and
[OI]$\lambda$6300/[OIII]$\lambda$5007$>$1/3
(and lower for `[OI]-weak' LINERS),
while NGC\,7589 shows
[OII]$\lambda$3727/[OIII]5007$\lambda$ $\approx$0.5
and [OI]$\lambda$6300/[OIII]$\lambda$5007 = 0.1.  
This places NGC\,7589 in a region populated by Seyferts
(see Fig.\,1b of Filippenko \& Terlevich 1992) but, again, close
to the borderline with LINERs.
We note that the measured line ratios are 
insensitive to the galaxy spectrum subtraction---the same 
result holds even the unsubtracted spectrum was used.
The detection of a broad component in H$\alpha$
reveals a type-I AGN and leads
to the classification of \ngc as either a Seyfert\,1.9
or possibly a LINER\,I.

\section{Discussion}
\label{sect:disc}
\subsection{Classification of  NGC\,7589}
\label{sect:discu_class}
On the classification diagrams based on emission line ratios, 
the distribution of the LLAGNs in the Ho et al.\ (2004) sample
has a continuous and smooth transition between Seyferts and LINERs.
This makes the conventional, `operational' dividing lines 
somewhat arbitrary and physically insignificant.
\ngc falls within the area in which galaxies are classified as Seyferts,
however, it is too close to the border region to 
allow an ultimate classification as either a Seyfert or LINER.
A similar conclusion holds when the
emission line width is considered (e.g.\ Ho et al.\ 2003).
We therefore treat \ngc as a member
of the important `border-line objects',
which probe the physics in transition from Seyfert to LINER.
Optically \ngc is a giant, bulge-dominated low surface brightness galaxy
(Pickering et al.\ 1997).

The high-amplitude X-ray variability detected is exceptional,
given the transition nature of \ngce.
While there is a trend in Seyferts showing
enhanced X-ray variability with decreasing luminosity
(e.g.\ Nandra et al.\ 1997), the trend breaks at low luminosities;
in particular LINERs do not vary much in X-rays 
(e.g.\ Ptak et al.\ 1998, Komossa et al.\ 1999,
Terashima et al.\ 2002). 
Those that do rarely exceed a factor of two variability.
The high-amplitude variation in the X-ray luminosity of
the order of magnitudes of $10^{42}$\,\ulum on timescales  of a few years
suggests black hole accretion as the ultimate source of the power in \ngc. 
This is in line with the AGN nature of \ngc indicated by the broad
$H\alpha$ line.

We may ask whether
the XMM `high-state' or the ROSAT `low-state' 
is more representative of the average X-ray luminosity.
We assume that the narrow line luminosities in the SDSS spectrum 
represents the average state of \ngce.
By making use of the $L_{\rm H\alpha}$$\sim$$L_{\rm x}$ correlation
for LLAGN (Halderson et al.\ 2001),
a luminosity of  $L_{\rm x}$$\sim$1\,$10^{41}$\,\ulum
($\sim$1.5\,$10^{41}$)  is predicted 
in the 0.1--2.4\,keV (0.5--10keV) band assuming $\Gamma$=\,1.7.
This value is significantly below the high-state luminosity, 
and is comparable to the low-state upper limit.
This suggests that probably  \ngc normally behaves  like a LLAGN,
and underwent an outburst that was caught by XMM.
At the high-state, $L_{\rm x}$
reached the lower end of typical Seyfert
luminosity range ($10^{42-44}$\,\ulume),
and the very upper end of LINERs.
The power-law model is again consistent with both
Seyfert and LINER types,
while the partial covering/reflection model would favor a Seyfert.

\subsection{The X-ray variability}

In general, 
X-ray variability in AGN  
may have two distinct 
origins: variations in intrinsic radiation and variations
in absorption/obscuration.
Below we discuss briefly our results in this context.
For \ngce, the mass of the central black hole is of 
the order of $10^{7}$\,\msune,
as found in our on-going follow-up work (in preparation).

{\em Variable absorption}:
Evidence for variable absorption has been found in some 
Seyfert galaxies, especially in type\,2 and intermediate-type, 
on timescales of months to years (e.g., Risaliti et al. 2002)
through variation of \nh and/or covering factor 
(e.g.\ NGC 3227: Komossa et al.\ 2001, NGC 3516: Guainazzi et al.\ 2001).
The absorbers are postulated to be gas clouds in the `broad-line region' 
(BLR) producing relatively fast variations, and
a parsec-scale `torus' producing  relatively long-term variations.
In particular, \ngc might be an analogue to  the LLAGN in M\,51, 
in which a suspected large-amplitude X-ray variation could be well 
explained by placing a thick (\nhe$>10^{24}$) absorber on and off 
the line-of-sight (Fukazawa et al.\ 2001).

For the ROSAT low-state,
taking the best-fit {\em cold} partial covering model for the
XMM spectrum at the highest state (Table\,\ref{tab:spec_fit})
and increasing artificially the covering factor up to $\ga0.95$,
the soft X-ray flux was then suppressed 
below the ROSAT detection limit.
For the  flux decrease between the two XMM observations,
we considered the joint fit
with the partial covering model (Table\,\ref{tab:spec_fit})
by assuming the continuum levels  to be the same and tieing them
together, and we still obtained an acceptable fit,
though slightly worse than that in Table\,\ref{tab:spec_fit};
this yielded a much larger value of \nh and covering factor
(8.0\,$10^{23}$\,\unh and 0.72) at the epoch of orbit 361 than 
those of orbit 272 (1.9\,$10^{23}$\,\unh and 0.29).
Both the timescales of $<$5\,years and of 5\,months are consistent
with the motion of gas clouds ($\approx$3,400\,km\,s$^{-1}$)
in the BLR for a $\sim10^{7}$\,\msun black hole.

{\em Variable intrinsic emission}:
For the ROSAT low-state, 
the Eddington ratio was estimated conservatively to be
\eddrate$<$ several times $10^{-4}$, 
much lower than the critical accretion rate (see Sect.\,1). 
This value indicates that the accretion proceeded via a RIAF.
A RIAF has been postulated to be present in some LLAGN
(e.g.\ Fabian \& Rees 1995, 
Lasota et al.\ 1996, Quataert et al.\ 1999, Ptak et al.\ 2004).
In particular, the proto-type of such a class,  the LINER NGC\,4258
(Lasota et al.\ 1996), has values of 
$M_{\rm bh}$ and $L_{\rm x}$ 
similar to those we found for \ngce.
A thin disc would be truncated and only exist at large radii 
(e.g.\ Quataert et al.\ 1999, Lu \& Wang 2000,
Meyer \& Meyer-Hofmeister 2002).
The broad iron line and reflection hump, which are characteristic of
typical Seyferts and believed to be evidence for a thin disc, 
are not expected to be seen in a RIAF.
These different accretion modes  also predict distinct spectral shapes
in the broad band spectral energy distribution (SED) and 
possibly in the X-ray spectra.
Therefore a good X-ray spectrum and/or SED
in a low-state is essential to test the RIAF hypothesis.

In the XMM high-state,
the identification of the accretion mode is not clear.
The  \eddrat was estimated to be a few per cent,
which is immediately close to the critical value.
Thus the accretion flow was possibly in a transition state between 
a RIAF and a thin disc.
Had the flux ever reached a peak much higher than 
the XMM high-state sometime between the observations,
a transition of the accretion mode might have taken place,
according to the current accretion theories.
The observed timescale of $\la$5 years 
is consistent with the mass diffusion timescale on which 
a thin disc drifted inwards down to several $R_{\rm s}$ 
from a transition radius of up to $\sim$\,hundreds $R_{\rm s}$. 
However, the quality of the obtained X-ray spectra does not allow
to distinguish a dominating thin disc 
(extending down to a few Schwarzschild radii $R_{\rm s}$) 
and a disc truncated at $\sim$\,hundreds $R_{\rm s}$,
though the probable spectral flattening---if this is indeed due to
reflection of the X-rays, would favour the former.
Deep X-ray observations for \ngc and of other objects of this kind
are essential to confront seriously current accretion theories.

Large (by a factor of 10 or higher) 
X-ray variability events appear to be rare.
The first results from our programme, which  searched 
over 386 XMM fields  previously covered with ROSAT pointed observations, 
yielded only one, or possibily two, detections.
The total sky area covered (66 sqr.\ deg.) contains, statisically,
$\sim$26 galaxies brighter than 15\,mag$_B$ and $\sim$180 galaxies
brighter than 16.4\,mag$_B$ (cf.\ \ngc of 15.2\,mag$_B$). 
A further extensive search, as well as the measurement of
the variability timescale,  is needed in order to 
be able to estimate the  event rate.

\section*{Acknowledgements}
We thank Cheng Li and Tinggui Wang for helping with the subtraction of
the stellar spectrum, and Christopher Brindle for help 
in verifying the XMM OM results.
W.Y.\ thanks Matteo Guainazzi for useful discussion.
Funding for the creation and distribution of the SDSS Archive has
been provided by the Alfred P.\ Sloan Foundation, the Participating
     Institutions, the National Aeronautics 
and Space Administration, the
      National Science Foundation, the U.S.\ 
Department of Energy, the
 Japanese Monbukagakusho, and the Max Planck Society.
This research has made use of the NASA/IPAC Extragalactic Database (NED).

\end{document}